\documentclass[12pt]{iopart}
\usepackage{iopams}  
\usepackage[noadjust]{cite}
\usepackage{footmisc}
\begin{document}

\title{Contextuality is About Identity of Random Variables}

\author{Ehtibar N. Dzhafarov\textsuperscript{1} and Janne V. Kujala\textsuperscript{2}}

\address{ \textsuperscript{1}Purdue University, USA, \textsuperscript{2}University
of Jyv\"askyl\"a, Finland}
\ead{ehtibar@purdue.edu, jvk@iki.fi}
\begin{abstract}
Contextual situations are those in which seemingly ``the same''
random variable changes its identity depending on the conditions under
which it is recorded. Such a change of identity is observed whenever
the assumption that the variable is one and the same under different
conditions leads to contradictions when one considers its joint distribution
with other random variables (this is the essence of all Bell-type
theorems). In our Contextuality-by-Default approach, instead of asking
why or how the conditions force ``one and the same'' random variable
to change ``its'' identity, any two random variables recorded under
different conditions are considered different ``automatically''.
They are never the same, nor are they jointly distributed, but one
can always impose on them a joint distribution (probabilistic coupling).
The special situations when there is a coupling in which these random
variables are equal with probability 1 are considered non-contextual.
Contextuality means that such couplings do not exist. We argue that
the determination of the identity of random variables by conditions
under which they are recorded is not a causal relationship and cannot violate laws of physics.\bigskip{} 


\end{abstract}

\vspace{2pc}
\maketitle

%\twocolumn

\section{Introduction}
\setcounter{footnote}{0}

The main purpose of this paper is to explain the principle of Contextuality-by-Default
(CbD) in nontechnical terms, and to demonstrate the conceptual clarity
 this principle brings in the analysis of random variables
recorded under varying conditions. The formulation of the principle
in our previous publications \cite{DK2014PLOS1,DK2014FoP,DK2013PLOS1,DKpressLNCS,DKpressADVCOG}
involves the following component principles:
\begin{description}
\item [{(Indexation-by-conditions)}] A random variable is \emph{identified}
(indexed, tagged) by all conditions under which its realizations are
recorded.
\item [{(Unrelatedness)}] Two or more random variables recorded under mutually
incompatible conditions are \emph{stochastically unrelated}, i.e.,
they possess no joint distribution.
\item [{(Coupling)}] A set of pairwise stochastically unrelated random
variables can be \emph{probabilistically coupled}, i.e., imposed a
joint distribution on; the choice of a coupling is generally non-unique.
\end{description}

CbD is complemented by the \emph{All-Posssible-Couplings} approach \cite{DK2014PLOS1,DK2014FoP,DK2013PLOS1,DKpressLNCS},
according to which the constraints satisfied by a set of random variables
observed under different conditions (e.g., the Tsirelson inequalities
\cite{TSIR1980} satisfied by spins in the EPR/Bohm paradigm \cite{BOHM})
can be characterized by studying the set of all possible ways in which
these random variables can be coupled. A general discussion of this
approach is left out of this paper. We focus, however, on the (im)possibility
of imposing on a set of random variables special,\emph{ identity couplings,
}representing situations considered \emph{non-contextual}. An identity
coupling is one in which random variables recorded under different
conditions are equal to each other with probability one.
This approach to (non-)contextuality was first explored by Larsson \cite{2Larsson}.\footnote{Non-contextuality can be generalized to be represented by couplings
that are as close to identity couplings as it is allowed by the marginal
distributions of the variables involved \cite{new,newlong}. This allows one
to extend the notion of contextuality to signaling systems. We leave
this (very recent) development outside the scope of this paper.} 

CbD is squarely within the framework of the Kolmogorovian probability
theory (KPT), although to keep the presentation nontechnical, we avoid
using here explicit measure-theoretic formalisms (cf.\ \cite{DK2010Front,DK2012LNCS,DK2013PAMS} and, especially,
\cite{DKarxivNHMP}). Our position is that there are no empirical
or theoretical considerations in quantum mechanics, cognitive science,
or anywhere else, that involve random variables but cannot be fully
described in the language of KPT. CbD can in fact be viewed as a principle
that ensures the universality of the descriptive power of KPT. It
is another matter (not elaborated in this paper, cf.\ \cite{DK2014FoP}) that KPT may not be
the most economic, convenient, or useful language for describing quantum
phenomena.

\section{Indexation-by-conditions}

The term ``conditions" for a random variable $C$ refers to any variable
$\gamma$ whose values (random, predictable, or controllable at will)
are paired with the observed realizations of these random variables.
The most familiar pairing employed in empirical sciences is chronological:
the values of $\gamma$ and the values of $C$ are recorded at the
same time, say, in a series of observations. 

Consider a toy example. Pat has a monitor that at any given
time shows a pair of symbols, $00,01,10,$ or $11$. The pairs follow
each other in a very long sequence, such as
\begin{equation}
01,11,11,10,10,10,11,11,01,\ldots\label{eq:sequence}
\end{equation}
Assume, for simplicity, that Pat is unable to record their order,\footnote{Enumerating observed realizations of a random variable amounts to introducing the ordinal number of the observation as a special condition under which the random variable is recorded. This presents no conceptual difficulties but complicates the discussion.}
so she simply counts the occurrences of each of $00,01,10,11$. Pat
wishes to treat these pairs as four possible values of a random variable
$C$, with a distribution 
\begin{equation}
\begin{array}{ccccccc}
00 &  & 01 &  & 10 &  & 11\\
p_{00} &  & p_{01} &  & p_{10} &  & p_{11}
\end{array}.
\end{equation}
So far, $\gamma$ is an empty notion: it can be viewed as a variable
having one and the same single value for each realization of $C$. 

Assume now that Pat notices that the pairs $00,01,10,11$ representing
$C$ are sometimes shown in red color and sometimes in blue. The two
colors may alternate randomly or in some regular fashion, e.g., red-blue-red-blue-$\ldots$.
In either case, Pat is able to count the occurrences of the four different
$C$ values separately for the blue and for the red colors, and form thereby
two random variables

\begin{equation}
\begin{array}{c}
C_{red}\sim\begin{array}{ccccccc}
00 &  & 01 &  & 10 &  & 11\\
p_{00} &  & p_{01} &  & p_{10} &  & p_{11}
\end{array},\\
\\
C_{blue}\sim\begin{array}{ccccccc}
00 &  & 01 &  & 10 &  & 11\\
p'_{00} &  & p'_{01} &  & p'_{10} &  & p'_{11}
\end{array},
\end{array}\label{eq:red-blue}
\end{equation}
where $\sim$ stands for ``is distributed as''. In this representation,
the two random variables have the same possible values, but are distinguished
by the condition $\gamma$ having two values, ``blue'' and ``red''.
That they are two different random variables is obvious if $\left(p_{00},p_{01},p_{10},p_{11}\right)\not=\left(p'_{00},p'_{01},p'_{10},p'_{11}\right)$,
but this is true even if the the two distributions in (\ref{eq:red-blue})
are the same. One way to see this is to observe that these distributions can always
be made different by viewing the conditions under which the variable
was recorded as part of its value. That is, we can have 
\begin{equation}
\begin{array}{c}
C_{red}\sim\begin{array}{ccccccc}
red\:00 &  & red\:01 &  & red\:10 &  & red\:11\\
p_{00} &  & p_{01} &  & p_{10} &  & p_{11}
\end{array},\\
\\
C_{blue}\sim\begin{array}{ccccccc}
blue\:00 &  & blue\:01 &  & blue\:10 &  & blue\:11\\
p'_{00} &  & p'_{01} &  & p'_{10} &  & p'_{11}
\end{array},
\end{array}\label{eq:red-blue values}
\end{equation}
where each value has the structure ``color$\:ij$". Even if $\left(p_{00},p_{01},p_{10},p_{11}\right)=\left(p'_{00},p'_{01},p'_{10},p'_{11}\right)$,
the two random variables have different distributions, simply because
they have different possible values. Such a redefinition is always
possible, and even when it is not convenient and one uses (\ref{eq:red-blue})
instead, this consideration justifies accepting as a general principle
that \emph{different conditions always define different random variables}. 

Obviously, the color of the symbols in this example can be replaced
by any condition that can be systematically associated with the recorded
values in sequence (\ref{eq:sequence}). Thus, Pat could simply distinguish
odd-numbered and even-numbered presentations, or have her window sometimes
open and sometimes closed when observing the symbols.

\section{Joint Distributions and Stochastic Unrelatedness}

Suppose now that Pat wishes to treat $C$ as a vector consisting of
two random variables, 
\begin{equation}
\begin{array}{c}
A=f_{1}\left(C\right)=\textnormal{left-hand component of }C,\\
B=f_{2}\left(C\right)=\textnormal{right-hand component of }C.
\end{array}\label{eq: projections}
\end{equation}
Being functions of one and the same random variable $C$, the random
variables $A$ and $B$ are \emph{jointly distributed}, i.e., for
every pair of values $A=i$ and $B=j$, Pat can uniquely determine
the probability with which these two values \emph{co-occur}. In this
case,
\begin{equation}
\begin{array}{c}
\Pr\left[A=i\textnormal{ and }B=j\right]=\Pr\left[C=ij\right],\\
i,j\in\left\{ 0,1\right\} .
\end{array}\label{eq: C=00003Dij}
\end{equation}
The co-occurrence in this example is chronological:
$i$ and $j$ occur simultaneously, within a single pair displayed on Pat's monitor. But the
deeper, more general meaning of the co-occurrence is that 
\begin{enumerate}
\item there are function $A=f_{1}\left(C\right)$ and $B=f_{2}\left(C\right)$
of one and the same random variable $C$; and 
\item the co-occurring values are $i=f_{1}\left(c\right)$ and $j=f_{2}\left(c\right)$
for any one value $c$ of $C$. 
\end{enumerate}
The joint distribution of $A$ and $B$ in any such case is uniquely
determined: for any value $\left(i,j\right)$ of $\left(A,B\right)$
one determines the set $S_{ij}$ of the values $c$ of $C$ such that
$i=f_{1}\left(c\right)$ and $j=f_{2}\left(c\right)$, and one puts
\begin{equation}
\begin{array}{c}
\Pr\left[A=i\textnormal{ and }B=j\right]=\Pr\left[c\in S_{ij}\right],\\
i,j\in\left\{ 0,1\right\} .
\end{array}
\end{equation}
The reverse of this statement is also true: if the joint distribution
of $\left(A,B\right)$ is well-defined, then $A$ and $B$ can be
presented as functions of one and the same random variable $C$. To
prove this, put $C=\left(A,B\right)$ with $\Pr\left[C=\left(i,j\right)\right]$
defined as $\Pr\left[A=i\textnormal{ and }B=j\right]$. This amounts
to using (\ref{eq: C=00003Dij}) as the definition of $C$. Then $A$
and $B$ as functions of $C$ are defined by (\ref{eq: projections}).

This can be generalized to any set of random variables of arbitrary
nature: for any such a set, the random variables comprising it have a joint
distribution if and only if they can be presented as functions of
one and the same random variable \cite{DK2014FoP,DK2010Front,DK2012LNCS,DKarxivNHMP}.
(Note that any set of jointly distributed random variables is a random
variable: the difference between a ``single'' random variable and,
say, a vector of several random variables is entirely superficial,
and can always be eliminated by renaming of the values, e.g., $\left(0,0\right),\left(0,1\right),\left(1,0\right),\left(1,1\right)$
into $1,2,3,4$.)

The situation is very different when we consider random variables
recorded under mutually incompatible conditions (i.e., different values
of $\gamma$). Thus, unlike $A$ and $B$ in the above example, the
variables $C_{red}$ and $C_{blue}$ never co-occur in the chronological
sense, and Pat would not know how to assign probability values to
the logically possible pairs
\begin{equation}
\begin{array}{c}
C_{red}=x\textnormal{ and }C_{blue}=y,\\
x,y\in\left\{ 00,01,10,11\right\} .
\end{array}\label{eq: xy pairs}
\end{equation}
Pat is not able to assess this probability by counting the occurrences
of the different pairs, because she does not know which value of
$C_{red}$ she should pair with which value of $C_{blue}$ to form
an ``observed'' value of the hypothetical random variable $C'=\left(C_{red},C_{blue}\right)$.
This situation is described by saying that $C_{red}$ and $C_{blue}$
are \emph{stochastically unrelated}. In view of what was said above, it
means that $C_{red}$ and $C_{blue}$ are not functions of any single
random variable. 

This does not mean, however, that $C_{red}$ and $C_{blue}$ cannot
be\emph{ imposed} a joint distribution on. The precise meaning of this is
as follows.

\section{Couplings }

Suppose that Pat formed the pairs $\left(x,y\right)$ in (\ref{eq: xy pairs})
according to some arbitrarily chosen scheme. Any such a pairing scheme
defines a (probabilistic) \emph{coupling} for $C_{red}$ and $C_{blue}$
\cite{TORR}. Formally, a coupling for $C_{red}$ and $C_{blue}$
is a random variable $Z=\left(X,Y\right)$ that satisfies
\begin{equation}
X\sim C_{red},\ Y\sim C_{blue}.\label{eq: coupling for CC}
\end{equation}
Calling $Z=\left(X,Y\right)$ a random variable means that $X$ and
$Y$ are jointly distributed. $Z$ is a coupling because the marginal
distributions of $X$ and $Y$, taken separately, are the same as
those of, respectively, $C_{red}$ and $C_{blue}$ in (\ref{eq:red-blue})
or (\ref{eq:red-blue values}). Note that $Z$ is a random variables different from both
$C_{red}$ and $C_{blue}$ (and in fact stochastically unrelated to them): by constructing 
a $Z$ satisfying (\ref{eq: coupling for CC}) one does not make $C_{red}$ and $C_{blue}$ jointly distributed
or changed in any way.

The coupling $Z$ is generally non-unique. In the matrix below,
\begin{equation}
\begin{tabular}{|c|cccc|c|}
\multicolumn{1}{c}{} &  &  &  & \multicolumn{1}{c}{} & \multicolumn{1}{c}{}\tabularnewline
\cline{2-6} 
\multicolumn{1}{c|}{} & \multicolumn{1}{c|}{$Y=00$} & \multicolumn{1}{c|}{$Y=01$} & \multicolumn{1}{c|}{$Y=10$} & $Y=11$ & \tabularnewline
\hline 
$X=00$ & \multicolumn{1}{c|}{$p_{0000}$} &  &  &  & $p_{00}$\tabularnewline
\cline{1-2} \cline{6-6} 
$X=01$ &  & $\ldots$ &  &  & $p_{01}$\tabularnewline
\cline{1-1} \cline{6-6} 
$X=10$ &  &  & $\ldots$ &  & $p_{10}$\tabularnewline
\cline{1-1} \cline{5-6} 
$X=11$ &  &  & \multicolumn{1}{c|}{} & $p_{1111}$ & $p_{11}$\tabularnewline
\hline 
 & \multicolumn{1}{c|}{$p'_{00}$} & \multicolumn{1}{c|}{$p'_{01}$} & \multicolumn{1}{c|}{$p'_{10}$} & $p'_{11}$ & \multicolumn{1}{c}{}\tabularnewline
\cline{1-5} 
\end{tabular}
\end{equation}
\bigskip
%
%\begin{center}
%Table 1
%\par\end{center}
%
%\noindent 
any of the (generally infinite) fillings of the interior that agrees with the indicated
marginal probabilities will define a possible coupling. The agreement
with the marginal probabilities means 
\begin{equation}
\begin{array}{c}
p_{x00}+p_{x01}+p_{x10}+p_{x11}=p_{x},\\
p_{00y}+p_{01y}+p_{10y}+p_{11y}=p'_{y},\\
x,y\in\left\{ 00,01,10,11\right\} ,
\end{array}\label{eq: coupling for CC probs}
\end{equation}
which is merely an explicit version of (\ref{eq: coupling for CC}).
For instance, Pat can form $Z=\left(X,Y\right)$ in such a way that
\begin{equation}
\begin{array}{c}
\Pr\left[X=x\textnormal{ and }Y=y\right]=\Pr\left[X=x\right]\Pr\left[Y=y\right]=p_{x}p'_{y},\\
x,y\in\left\{ 00,01,10,11\right\} .
\end{array}
\end{equation}
This $Z$ is called an \emph{independent coupling},
and it is universally imposable on any set of pairwise stochastically
unrelated random variables (which is the reason stochastic unrelatedness
is often confused with stochastic independence, which is a form of
stochastic relationship). 

Equations (\ref{eq: coupling for CC})-(\ref{eq: coupling for CC probs}), however,
rule out certain subclasses of couplings. Thus, Pat may be especially
interested in whether she can simply treat $C_{red}$ and $C_{blue}$
as ``essentially one and the same'' random variable. The rigorous meaning
of ``essentially the same'' is the \emph{identity coupling}, defined
by the conjunction of (\ref{eq: coupling for CC}) and (\ref{eq: coupling for CC probs})
with
\begin{equation}
\Pr\left[X=Y\right]=1,
\end{equation}
or, if Pat uses (\ref{eq:red-blue values}) instead of (\ref{eq:red-blue}),
\begin{equation}
  \Pr\left[
    \begin{tabular}{c}
      \(X=red\:x\textnormal{ and }Y=blue\:x\)\\
      for some \(x\in\left\{ 00,01,10,11\right\}\)
    \end{tabular}
    \right]=1,
\end{equation}
Obviously this identity coupling exists if and only if $p_{x}=p'_{x}$
for all $x$, i.e., if and only if $C_{red}$ and $C_{blue}$ in representation
(\ref{eq:red-blue}) have the same distribution.

\section{Same Identity vs Same Distributions}

Being identically distributed, however, does not generally guarantee
the possibility of the identity coupling. To see this, let us assume
that Pat views $C$ as a pair $\left(A,B\right)$ defined by (\ref{eq: projections}).
In accordance with the indexing-by-conditions principle, she has then
$C_{red}=\left(A_{red},B_{red}\right)$ and $C_{blue}=\left(A_{blue},B_{blue}\right)$,
i.e., both $A$ and $B$, since they are recorded in conjunction with
$\gamma=\textnormal{red/blue}$, are to be indexed by these conditions.
The distributions of $C_{red}$ and $C_{blue}$ are then represented
by two joint distributions,
\begin{equation}
\begin{tabular}{|c|c|c|c}
\multicolumn{1}{c}{} & \multicolumn{1}{c}{} & \multicolumn{1}{c}{} & \tabularnewline
\cline{1-3} 
$\gamma=\textnormal{red}$ & $B_{red}=0$ & $B_{red}=1$ & \tabularnewline
\hline 
$A_{red}=0$ & $p_{00}$ & $p_{01}$ & \multicolumn{1}{c|}{$p_{0\cdot}$}\tabularnewline
\hline 
$A_{red}=1$ & $p_{10}$ & $p_{11}$ & \multicolumn{1}{c|}{$p_{1\cdot}$}\tabularnewline
\hline 
\multicolumn{1}{c|}{} & $p_{\cdot0}$ & $p_{\cdot1}$ & \tabularnewline
\cline{2-3} 
%\end{tabular}
%\end{equation}
%
%\begin{equation}
%\begin{tabular}{|c|c|c|c}
\multicolumn{1}{c}{} & \multicolumn{1}{c}{} & \multicolumn{1}{c}{} & \tabularnewline
\cline{1-3} 
$\gamma=\textnormal{blue}$ & $B_{blue}=0$ & $B_{blue}=1$ & \tabularnewline
\hline 
$A_{blue}=0$ & $p'_{00}$ & $p'_{01}$ & \multicolumn{1}{c|}{$p'_{0\cdot}$}\tabularnewline
\hline 
$B_{blue}=1$ & $p'_{10}$ & $p'_{11}$ & \multicolumn{1}{c|}{$p'_{1\cdot}$}\tabularnewline
\hline 
\multicolumn{1}{c|}{} & $p'_{\cdot0}$ & $p'_{\cdot1}$ & \tabularnewline
\cline{2-3} 
\end{tabular}
\end{equation}
\bigskip
%
%\begin{center}
%Table 2
%\par\end{center}
%

Suppose first that Pat is only interested in whether she can treat
$A_{red}$ and $A_{blue}$ as an ``essentially the same'' random variable
(disregarding $B$). This translates into the question of the existence
of the identity coupling for $A_{red}$ and $A_{blue}$, i.e., a random
variable $\left(X,X'\right)$ with
\begin{equation}
  X\sim A_{red},\ X'\sim A_{blue},\textnormal{ and }\Pr\left[X=X'\right]=1.
\end{equation}
Repeating the reasoning of the previous subsection, Pat comes to the
conclusion that such a coupling exists if and only if $A_{red}$ and
$A_{blue}$ are identically distributed, i.e., $p_{0\cdot}=p'_{0\cdot}$.
The situation is analogous for $B_{red}$ and $B_{blue}$: the identity
coupling $\left(Y,Y'\right)$ for them exists if an only if $B_{red}\sim B_{blue}$,
i.e., $p_{\cdot0}=p'_{\cdot0}$. 

Let now both these conditions be satisfied: $A_{red}\sim A_{blue}$
and $B_{red}\sim B_{blue}$, i.e., let Pat deal with the distributions
\begin{equation}\label{table3}
\begin{tabular}{|c|c|c|c}
\multicolumn{1}{c}{} & \multicolumn{1}{c}{} & \multicolumn{1}{c}{} & \tabularnewline
\cline{1-3} 
$\gamma=\textnormal{red}$ & $B_{red}=0$ & $B_{red}=1$ & \tabularnewline
\hline 
$A_{red}=0$ & $p_{00}$ & $p_{01}$ & \multicolumn{1}{c|}{$p_{0\cdot}$}\tabularnewline
\hline 
$A_{red}=1$ & $p_{10}$ & $p_{11}$ & \multicolumn{1}{c|}{$p_{1\cdot}$}\tabularnewline
\hline 
\multicolumn{1}{c|}{} & $p_{\cdot0}$ & $p_{\cdot1}$ & \tabularnewline
\cline{2-3} 
%\end{tabular}
%\end{equation}
%
%\begin{equation}
%\begin{tabular}{|c|c|c|c}
\multicolumn{1}{c}{} & \multicolumn{1}{c}{} & \multicolumn{1}{c}{} & \tabularnewline
\cline{1-3} 
$\gamma=\textnormal{blue}$ & $B_{blue}=0$ & $B_{blue}=1$ & \tabularnewline
\hline 
$A_{blue}=0$ & $p'_{00}$ & $p'_{01}$ & \multicolumn{1}{c|}{$p_{0\cdot}$}\tabularnewline
\hline 
$B_{blue}=1$ & $p'_{10}$ & $p'_{11}$ & \multicolumn{1}{c|}{$p_{1\cdot}$}\tabularnewline
\hline 
\multicolumn{1}{c|}{} & $p_{\cdot0}$ & $p_{\cdot1}$ & \tabularnewline
\cline{2-3} 
\end{tabular}
\end{equation}
\bigskip
%
%\begin{center}
%Table 3
%\par\end{center}
%
Any random variable $Z=\left(X,Y,X',Y'\right)$ such that
\begin{equation}
(X,Y)\sim (A_{red},B_{red}),\ (X',Y')\sim (A_{blue},B_{blue})
\label{eq: distributional}
\end{equation}
is a coupling for $\left(A_{red},B_{red}\right)$ and $\left(A_{blue},B_{blue}\right)$.
It is easy to see that $Z$ is also a coupling for separately taken $A_{red},B_{red},A_{blue},B_{blue}$,
because (\ref{eq: distributional}) implies
\begin{equation}
X\sim A_{red},\ X'\sim A_{blue},\ Y\sim B_{red},\ Y'\sim B_{blue}.
\end{equation}
Let now the question Pat poses for herself be whether the red/blue
difference matters when considering \emph{both} $A$ and $B$ together. This
questions translates into that of the possibility of $Z$ being an
identity coupling satisfying
\begin{equation}
\begin{tabular}{c}
\(\Pr\left[X=X'\right]=1\),\\
\(\Pr\left[Y=Y'\right]=1\).
\end{tabular}
\label{eq: coupling for both}
\end{equation}
Even though $A_{red}\sim A_{blue}$ and $B_{red}\sim B_{blue}$, such a coupling may not exist.
It is clear that it exists (generally non-uniquely) if and only
if the two joint distributions are identical, which in this case (with
$p_{0\cdot}$ and $p_{\cdot0}$ fixed) is equivalent to $p_{00}=p'_{00}$.
If $p_{00}\not=p'_{00}$, then, in any coupling, one or both of the
equations in (\ref{eq: coupling for both}) should be violated. 

This leads us to the notion of probabilistic contextuality.

\section{Probabilistic Contextuality}

It can be said that when $\left(A_{red},B_{red}\right),\left(A_{blue},B_{blue}\right)$
cannot be coupled by an identity coupling, the color creates a \emph{context}
for the probability distributions involved. It can be shown
that Pat can always find a value $p$ such that $Z=\left(X,Y,X',Y'\right)$
is a coupling for $\left(A_{red},B_{red}\right)$ and $\left(A_{blue},B_{blue}\right)$
that satisfies
\begin{equation}
\begin{array}{c}
\Pr\left[X=X'\right]=p,\\
\Pr\left[Y=Y'\right]=p.\\
\end{array}\label{eq: p<1}
\end{equation}
Choosing $p=1$ means having the identity coupling, and we take this
case as representing a \emph{lack of contextuality}. As mentioned
earlier, in the distributions described by (\ref{table3}), this is not the
case if $p_{00}\not=p'_{00}$. In this case $p$ should be chosen
to be less than 1. The minimum possible value of $1-p$ can in fact
be taken as a measure of \emph{contextuality}, i.e., a measure of
deviation of the system from the identity coupling representing lack
of contextuality. 

We will not, however, pursue the subject of quantitatively measuring
contextuality in this paper. We only want to establish the defining
aspect of contextuality: \vspace{1ex}

\noindent \emph{the contextuality in a system of random
variables recorded under various conditions is a deviation of the
possible couplings for this system from a specifically chosen identity coupling.} \vspace{1ex}

\noindent There
can be more than one identity coupling, depending on which of the
random variables involved are hypothesized to be ``essentially the
same'' despite being labeled by different conditions. To each specific choice
of an identity coupling there corresponds a specific meaning of contextuality.

Let us make this clear on the abstract notion of a system whose inputs
are $\alpha,\beta,\gamma,\ldots$ and whose outputs are $A,B,C\ldots$.
(This is not the most general conceptual set-up, but if one wants to
avoid technicalities, it is general enough.) The inputs are simply
variables, each having several possible values, while the outputs
are random variables with a well-defined joint distribution for each
possible combination of the inputs values. Let $\phi,\chi,\psi,\ldots$
be these possible combinations: we call them \emph{treatments} or
\emph{conditions}. By the indexation-by-conditions principle, the
outputs are to be labeled 
\begin{equation}
\left(A_{\phi},B_{\phi},C_{\phi},\ldots\right),\left(A_{\chi},B_{\chi},C_{\chi},\ldots\right),\left(A_{\psi},B_{\psi},C_{\psi},\ldots\right),\ldots,\label{eq: ABC exmaple}
\end{equation}
where any random variable is jointly distributed with any identically
indexed random variable but stochastically unrelated to any differently
indexed one. Any random variable 
\begin{equation}
U=\left(X_{\phi},Y_{\phi},Z_{\phi},\ldots,X_{\chi},Y_{\chi},Z_{\chi},\ldots,X_{\psi},Y_{\psi},Z_{\psi},\ldots\right)
\end{equation}
such that 
\begin{equation}
\begin{array}{c}
\left(X_{\phi},Y_{\phi},Z_{\phi}\ldots\right)\sim\left(A_{\phi},B_{\phi},C_{\phi},\ldots\right),\\
\left(X_{\chi},Y_{\chi},Z_{\chi},\ldots\right)\sim\left(A_{\chi},B_{\chi},C_{\chi},\ldots\right),\\
\left(X_{\psi},Y_{\psi},Z_{\psi},\ldots\right)\sim\left(A_{\psi},B_{\psi},C_{\psi},\ldots\right),\\
\ldots
\end{array}
\end{equation}
is a coupling for (\ref{eq: ABC exmaple}). 

Assume now that, for whatever reason, one thinks that of the inputs
$\alpha,\beta,\gamma,\ldots$ only $\alpha$ can influence the identity
of $A$ and only $\beta$ can influence the identity of $B$. This
means that if, e.g., $\phi\left(\alpha_{0}\right),\phi'\left(\alpha_{0}\right),\phi''\left(\alpha_{0}\right),\ldots$
denote treatments containing the same value $\alpha_{0}$ of $\alpha$, then all $A_{\phi\left(\alpha_{0}\right)},A_{\phi'\left(\alpha_{0}\right)},A_{\phi''\left(\alpha_{0}\right)},\ldots$
are ``essentially'' the same, even if differently labeled. A rigorous
formulation is that there exists a coupling $U'$ in which
\begin{equation}
\Pr\left[X_{\phi\left(\alpha_{0}\right)}=X_{\phi'\left(\alpha_{0}\right)}=X_{\phi''\left(\alpha_{0}\right)}=\ldots\right]=1,
\end{equation}
for every value $\alpha_{0}$ of $\alpha$. Analogously, the hypothesized
relation between $\beta$ and $B$ translates into the constraint
\begin{equation}
\left[Y_{\phi\left(\beta_{0}\right)}=Y_{\phi'\left(\beta_{0}\right)}=Y_{\phi''\left(\beta_{0}\right)}=\ldots\right]=1,
\end{equation}
for every value $\beta_{0}$ of $\beta$. In the sense of being subject
to these two sets of constraints, $U'$ is an identity coupling. If
now it can be shown that such a coupling does not exist, then the
system is contextual with respect to the identity coupling $U'$.
The interpretation is that the identity of $A$ depends not only on
$\alpha$ or/and the identity of $B$ depends not only on $\beta$.

\section{Alice-Bob EPR/Bohm Paradigm}\label{sec:Alice-Bob}

Let us now illustrate the notion of contextuality on an example well
familiar in quantum physics: two entangled spin-half particles, with
Alice and Bob measuring spins along two directions each. Let the direction
chosen by Alice be denoted $\alpha$, with values $\alpha_{1},\alpha_{2}$;
the direction chosen by Bob is denoted by $\beta$, with values $\beta_{1},\beta_{2}$.
We treat the four combinations $\left(\alpha_{i},\beta_{j}\right)$
($i,j\in\left\{ 1,2\right\} $) of settings by Alice and Bob as conditions
under which the spins are recorded, $A$ in Alice's particle, $B$
in Bob's, both random variables with possible values $\mbox{+1}$
and $-1$.

In accordance with the indexation-by-conditions and unrelatedness
principles, we have four stochastically unrelated to each other pairs
of random variables $\left(A_{ij},B_{ij}\right)$, $i,j\in\left\{ 1,2\right\} $.
They are distributed as

\begin{equation}
\begin{tabular}{|c|c|c|c}
\multicolumn{1}{c}{} & \multicolumn{1}{c}{} & \multicolumn{1}{c}{} & \tabularnewline
\cline{1-3} 
{$\alpha_{i},\beta_{j}$} & {$B_{ij}=+1$} & {$B_{ij}=-1$} & \tabularnewline
\hline 
{$A_{ij}=+1$} & {$p_{ij}$} & {$q_{ij}$} & \multicolumn{1}{c|}{{$p_{ij}+q_{ij}$}}\tabularnewline
\hline 
{$A_{ij}=-1$} & {$r_{ij}$} & {$s_{ij}$} & \tabularnewline
\cline{1-3} 
\multicolumn{1}{c|}{} & {$p_{ij}+r_{ij}$} & \multicolumn{1}{c}{} & \tabularnewline
\cline{2-2} 
\end{tabular}
\end{equation}
\bigskip
%
%\begin{center}
%Table 4
%\par\end{center}
%
%\noindent 
A coupling for these pairs of random variables is a random
variable
\begin{equation}
V=\left(X_{11},Y_{11},X_{12},Y_{12},X_{21},Y_{21},X_{22},Y_{22}\right)\label{eq: complete EPR}
\end{equation}
such that
\begin{equation}
\left(X_{ij},Y_{ij}\right)\sim\left(A_{ij},B_{ij}\right),\quad i,j\in\left\{ 1,2\right\}.
\label{eq: complete EPR req}
\end{equation}

It is taken as a given that a change in Alice's setting, $\alpha_{1}\rightarrow\alpha_{2}$,
changes the identity of Alice's random variable (and analogously for
Bob). This means that $X_{1j}$ and $X_{2j}$ in the coupling $V$
should not be required to be equal to each other, no matter what $j$
is (and analogously for $Y_{i1}$ and $Y_{i2}$). It seems, however,
reasonable to assume that Bob's settings ``have nothing to do" with Alice's
measurements, and vice versa. This translates into requiring that
\begin{equation}
\begin{array}{c}
\Pr\left[X_{i1}=X_{i2}\right]=1,\\
\Pr\left[Y_{1j}=Y_{2j}\right]=1,\\
i,j\in\left\{ 1,2\right\}.
\end{array}\label{eq: pre-reduced EPR}
\end{equation}
The coupling $V$ subject to this requirement can be chosen as the identity
coupling of special interest (this is only one of logically
possible identity couplings). Equivalently,
the requirement is that the four pairs $\left(A_{ij},B_{ij}\right)$
allow for a \emph{reduced coupling}
\begin{equation}
V'=\left(X'_{1},X'_{2},Y'_{1},Y'_{2}\right),\label{eq: reduced EPR}
\end{equation}
such that
\begin{equation}
\begin{array}{c}
\left(X'_{i},Y'_{j}\right)\sim\left(A_{ij},B_{ij}\right)\\
i,j\in\left\{ 1,2\right\} .
\end{array}\label{eq: reduced EPR req}
\end{equation}
This is the closest rigorous formulation for the usually considered
``joint distribution of $A_{1},A_{2},B_{1},B_{2}$'' \cite{FINE}.
We know that $V$ subject to (\ref{eq: pre-reduced EPR}) exists if
and only if the conjunction of the following two conditions is satisfied:
\begin{enumerate}
\item Marginal selectivity \cite{DKarxivNHMP,DK2012JMP,DK2014Topics} or
no-signaling \cite{CERE2000,MASANES},
\begin{equation}
\begin{array}{c}
p_{i1}+q_{i1}=p_{i2}+q_{i2}=p_{i\cdot},\\
p_{1j}+r_{1j}=p_{2j}+r_{2j}=p_{\cdot j},\\
i,j\in\left\{ 1,2\right\} ;
\end{array}
\end{equation}

\item CH/Fine inequalities \cite{FINE,CH74},
\begin{equation}
\begin{array}{c}
-1\leq p_{11}+p_{12}+p_{21}+p_{22}-\left(2p_{3-i,3-j}+p_{i\cdot}+p_{\cdot j}\right)\leq0,\\
i,j\in\left\{ 1,2\right\} .
\end{array}
\end{equation}

\end{enumerate}
We can say that if (and only if) these two requirements are jointly
met, then the system in question is non-contextual with respect to
the identity coupling $V$ defined by (\ref{eq: complete EPR})-(\ref{eq: complete EPR req})-(\ref{eq: pre-reduced EPR}),
or equivalently, (\ref{eq: reduced EPR})-(\ref{eq: reduced EPR req}).
The essence of all Bell-type theorems is to establish conditions for such non-contextuality.

Conversely, we can say that the system exhibits contextuality if and
only the two requirements are violated. It may be important or at least useful 
in many cases to distinguish the following two cases:
\begin{description}
\item[Case 1.] Marginal selectivity is violated, that is, either the distribution
of $B_{ij}$ (the spin recorded by Bob for the direction $\beta_{j}$
he chose) changes with $\alpha_{i}$ (the direction chosen by Alice),
or vice versa. This (perhaps) should be interpreted as a direct influence
of Bob's choices on Alice's measurements, i.e., some form of signaling.  Note that the way they are
written above, the probabilities $p_{i\cdot},p_{\cdot j}$ in CH/Fine
inequalities are not defined if marginal selectivity (no-signaling)
is violated. 
\item[Case 2.] Marginal selectivity is satisfied but CH/Fine inequalities are violated.
This can be called the case of ``pure contextuality''. Bob's settings
do not affect the distribution of Alice's recordings, they only determine
the way they are grouped into random variables (see the next section). The laws of quantum
physics and special relativity dictate this case when the two particles
are separated by a space-like interval. 
\end{description}
In all our previous publications regarding contextuality \cite{DK2014PLOS1,DK2014FoP,DK2013PLOS1,DKpressLNCS,DKpressADVCOG}
we only considered Case 2 as that of contextuality, preferring to
speak of ``direct cross-influences'' in Case 1. Intuitively, the
two cases must be distinguished, although perhaps not as sharply: 
direct cross-influences need not prevent the system from also being
contextual. There is in fact a principled way to define contextuality
``on top of'' violations of marginal selectivity \cite{new,newlong}.

\section{Why Contextual Indexing Does not Violate Laws of Physics}

Let us focus on Case 2 (``pure contextuality'') and ask ourselves:
if Bob's choices change the identity of Alice's measurements, does
not this mean a form of signaling from Bob to Alice? In the case of
a space-like separation this would contravene special relativity.
The answer is negative, and it can be justified on two levels. 

On a most obvious one, Alice can never guess that Bob even exists
if the only information she has is the distributions of spins in her
particle in response to her choices of settings. The non-identity
of $A_{i1}$ and $A_{i2}$ is not available to her. It can only be
established by someone else, a Charlie who receives the information
both about the settings and about the spin recording from both Alice
and Bob. From Charlie's point of view, Alice observes a mixture of
$A_{i1}$ and $A_{i2}$, but Alice cannot know that. 

On a deeper level, the negative answer is justified because the identity of a random variable
is not an objective, physical entity to begin with. The realizations
of random variables, such as ``spin-down in direction $\alpha_{1}$''
are objective, and the probabilities of all such observations (in the classical,
frequentist sense) among, say, all spins in direction $\alpha_1$ are objective too. 
But which realizations are grouped
together to count them and establish their relative frequencies is a matter of choice. 

It is analogous to looking at
a set of points scattered on a sheet of paper: one can group them
in this or that way and create various patterns without ever affecting
the objective locations of the points. For Charlie, $A_{ij}$ and
$B_{ij}$ are double-indexed because he chose to relate their realizations
to both Alice's and Bob's settings, $\alpha_{i}$ and $\beta_{j}$.
CbD prescribes doing this ``automatically'', because in this way
one can gain information (e.g., about quantum correlations between
entangled particles), and because in cases when this is redundant
one does not lose anything: in those cases differently labeled random
variables are simply merged within an identity coupling.

The latter point deserves being emphasized. The requirement to index
a random variable by all conditions paired with its realizations
may be interpreted as a call for some kind of ``all-is-one'' holism.
What if Charlie notices that the information he gets from Alice and Bob
is received either in
the morning, or during the daytime, or else in the evening? CbD requires then
from Charlie to index the spins as $A_{ijt}$ and $B_{ijt}$, where $t$ assumes
three values (morning, daytime, evening). Does not this ``automatic''
extra-labeling make the analysis unnecessarily complicated? The answer
is: Charlie can always choose not to record the time of the day, but
once he chooses to record it, he will either gain valuable knowledge
(if the time of the day turns out to affect the joint distributions
of the spins), or, in the worst case, he will find out that an identity coupling for $\left(A_{ijt},B_{ijt}\right)$
can be constructed eliminating the need for using $t$.

There is yet another, purely formal way of justifying why the identity
of random variables has no physical meaning. If we list all the random
variables in play, each indexed by the conditions under which it has
been recorded, their identities are entirely defined by (and define)
the coupling imposed on them. Different couplings correspond to different
identities. But couplings for one and the same system of random variables
are generally non-unique, and should therefore be viewed as no more
than possible (mutually exclusive) mathematical descriptions. The
totality of all couplings that are imposable on a given system does
characterize the system physically, as detailed in Refs. \cite{DK2013PLOS1,DK2014FoP,DK2014PLOS1,DKpressLNCS},
but no single coupling is more ``real'' than another.

\section{Concluding Remarks}

The position presented in this paper is summarized in the abstract,
and need not be repeated. 

Commenting on an earlier draft of this paper, Arkady Plotnitsky suggested a connection between CbD and Bohr's use of the notion of complementarity
in his 1935 reply \cite{Bohr} to the famous EPR paper \cite{EPR} (for a thorough analysis of this exchange, see Ref. \cite{PLOTN}).
Plotnitsky notes that Indexation-by-Conditions corresponds to Bohr\textquoteright{}s
view that each quantum phenomenon is unique, so that to specify it
one needs all the conditions under which it occurs; and that the Unrelatedness
and Coupling principles reflect Bohr's notion of complementary quantum
phenomena as being mutually exclusive but relatable to each other. 

Of the modern approaches to probabilistic contextuality in the literature on foundations
of quantum mechanics, Larsson's \cite{2Larsson} comes very close to ours, 
while Krennikov's approach \cite{2Khrennikov,3Khrennikov,3aKhrennikov} is in some respects more general. See also Refs. \cite{x1LAUD,x2SPEKK,x3KIRCH,x4BAD}. Here, we will briefly discuss two other treatments.

One of them is proposed by Avis, Fischer, Hilbert, and Khrennikov
\cite{Avis}. In Ref.~\cite{DK2014PLOS1} we called it ``\emph{conditionalization}" 
and compared it with CbD.
Conditionalization  consists in considering different values of $\gamma$
associated with realizations of a random variables as if $\gamma$
were a random variable in its own right. For instance, in our introductory
toy example, the color of Pat's random variable $C$ (with values
$00,01,10,11$) would be considered a random variable with two values,
``red'' and ``blue'', whether the color changes randomly or alternates
in some regular fashion, say, red-blue-red-blue-$\ldots$. The probabilities
assigned to the two values have to be nonzero; otherwise they are arbitrary
and play no role in characterizing the system being studied. The variable $C$ is then considered conditioned
on the values ``red'' and ``blue'', with the distributions in
(\ref{eq:red-blue}) treated as conditional distributions. This is
equivalent to the indexation-by-conditions in CbD: two different random variables are indexed by ``red'' and
``blue''. However, here the conditionalization analysis
ends, while the analysis led to by CbD approach only begins at this point: it entails considering
various couplings of the two variables and determining, when contextuality
is of the main interest, whether they contain identity couplings.

Another approach is based on the use of signed probability measures
(sometimes referred to as ``negative probabilities''). The approach
dates back to Paul Dirac \cite{Dirac}, but here it will be presented
primarily based on Refs.~\cite{OAS,ACACIO}. In relation to CbD, this
approach can be presented in terms as considering identity
couplings only, but allowing some of the joint probabilities in them
to be negative (and some greater than 1). As these are not then true couplings, they can be called \emph{quasi-couplings}.
This approach will not be applicable to our introductory toy example,
but it can be illustrated on the EPR/Bohm paradigm. Given $\left(A_{ij},B_{ij}\right)$,
$i,j\in\left\{ 1,2\right\} $, we construct a quasi-coupling $G=\left(E_{1},F_{1},E_{2},F_{2}\right)$
such that 
\begin{equation}
\begin{array}{c}
\Pr\left[E{}_{i}=a,F{}_{j}=b\right]=\Pr\left[A_{ij}=a,B_{ij}=a\right],\\
i,j\in\left\{ 1,2\right\} ,\; a,b\in\left\{ -1,+1\right\} .
\end{array}
\end{equation}
The quasi-coupling $G$, however, is not a conventional random variable,
in the sense that the signed probability values 
\begin{equation}
p\left(a_{1},b_{1},a_{2},b_{2}\right)=\Pr\left[E_{1}=a_{1},F_{1}=b_{1},E_{2}=a_{2},F_{2}=b_{2}\right]\label{eq: neg}
\end{equation}
while well-defined for all $a_{1},b_{1},a_{2},b_{2}\in\left\{ -1,+1\right\} $
and summing to 1, may be negative or greater than 1. If the Alice-Bob
system allows for the identity coupling in the sense of CbD, then
$G$ simply coincides with the corresponding reduced coupling $V'$ in (\ref{eq: reduced EPR}), with
all probabilities between $0$ and $1$. If the identity coupling
does not exist (i.e., we have a contextual system), then some of the
probabilities in (\ref{eq: neg}) will have to be negative. The relationship between CbD
and the signed probability measures is presently under investigation.
It should be noted that the approach in question is not applicable
if the marginal selectivity condition is violated (cf.\ the last paragraph of Section \ref{sec:Alice-Bob}). In fact the quasi-coupling
$G$ above exists if and only if marginal selectivity (no-signaling)
holds \cite{Sabri}.

\section*{Acknowledgments}

The authors are grateful to Andrei Khrennikov, Arkady Plotnitsky,
Samson Abramsky, Jan-\AA ke Larsson, Acacio de Barros, and Gary Oas for
many helpful discussions. The work was supported by NSF grant SES-1155956.

\end{document}